\documentstyle[emulateapj,epsfig]{article}

\received{2000 February 7}
\accepted{2000 April 20}
\slugcomment{Accepted 2000 April 20 for publication in The Astrophysical Journal}
\lefthead{Hanson, Still \& Fender}
\righthead{Cygnus X-3}

\begin{document}  

\title{
Orbital dynamics of Cygnus X-3
}
\author{M.M. Hanson$^1$, M.D. Still$^{2,4}$, R.P. Fender$^{3}$}
\affil{$^1$Department of Physics, The University of Cincinnati, Cincinnati, 
OH 45221-0011 \\ $^2$NASA/Goddard Space Flight Center, Code 662, Greenbelt, 
MD, 20771  \\ $^3$Astronomical Institute `Anton Pannekoek', University of Amsterdam,
and Center for High Energy Astrophysics, Kruislaan 403,
1098 SJ, Amsterdam, The Netherlands}

\altaffiltext{4}{Also Universities Space Research Association}

\begin{abstract}

Orbital-phased-resolved infrared spectra of Cygnus X-3
in outburst and quiescence, including tomographic
analysis, are presented.  We confirm the phasing of
broad He~{\sc ii} and N~{\sc v} lines in quiescence,
such that maximum blue shift corresponds to the X-ray
minimum at $\Phi = 0.00 \pm 0.04$.  In outburst,
double-peaked He~{\sc i} structures show a similar
phasing with two significant differences: (a) although
varying in relative strength, there is continuous line
emission in blue and red peaks around the orbit, and (b)
an absorption component, $\sim 1/4$ of an orbit out of
phase with the emission features, is discerned.  
Doppler tomograms of
the double-peaked profiles are consistent with a
disk-wind geometry, rotating at velocities of 1000 km
s$^{-1}$.  Regrettably, the tomography algorithm will
produce a similar ring structure from alternative line
sources if contaminated by overlying P Cygni profiles.
This is certainly the case in the strong 2.0587~$\mu$m
He~{\sc i} line, leading to an ambiguous solution for
the nature of double-peaked emission.  The absorption 
feature, detected 1/4 of an orbit out of phase with the 
emission features, is consistent with an origin in
the He star wind and yields for the first time a plausible radial
velocity curve for the system.  We directly derive the
mass function of the system, 0.027~M$_{\odot}$.  If we
assume a neutron star accretor and adopt a high orbital
inclination, $i > 60^{\circ}$, we obtain a mass range for the 
He star of 5~M$_{\odot}$ $\la$ M$_{\mbox{\tiny WR}}$ $\la$
11~M$_{\odot}$.  Alternatively if the compact object is
a black hole, we estimate M$_{\mbox{\tiny BH}}$ $\la$ 10
M$_{\odot}$.  We discuss the implications of these
masses for the nature and size of the binary system.

\end{abstract} 


\section{INTRODUCTION}  

First discovered in the pioneering  X-ray surveys of the 1960s, Cygnus
X-3 (Cyg X-3) remains among the least  understood of the long-standing
X-ray sources known in the sky.  In the early 1970s, it was discovered
to show a flux modulation  on a period of 4.8  hours, first in  X-rays
(Parsignault et  al.\  1972; Sanford \& Hawkins  1972) and then also at
near-infrared wavelengths (Becklin et al.\  1973), which was assumed to
be the orbital period  of the binary system.  Modulation of the  light
curve at this period is asymmetric and extremely similar in both X-ray
and infrared bands (Mason, Cordova \&  White 1986), with the exception
of irregular, rapid flaring superposed   on the orbital modulation  in
the infrared  (Mason et al.\ 1986; Fender  et al.\ 1996).  The period of
this orbit  was  further  shown  to   be  rapidly increasing, with   a
time-scale of just 850,000 yrs (Kitamoto et al.\ 1995).

Van den Heuvel \& De Loore (1973) made  the prediction that the system
was comprised of a compact object and a helium star, thus representing
a remarkably rare, late evolutionary  stage of a massive X-ray binary.
Nearly 20 years later, persuasive evidence  to support this prediction
was  found  in  near-infrared  spectroscopic  data  presented  by  van
Kerkwijk et al.\ (1992;  1996).  They discovered strong,  broad emission
lines of  neutral and ionized helium, combined  with a  marked lack of
hydrogen features. These properties are reminiscent of massive, highly
evolved stars known as Wolf-Rayet stars (Smith 1968).  As such, Cyg X-3
represents  one  of  the  most unique binary   systems in  our Galaxy.
Should  the  helium companion star become a supernova, as is generally
predicted  for Wolf-Rayet stars (however,  cf.  Conti 1996), it could
evolve  into  a  double compact  object   binary system, such   as the
Hulse-Taylor pulsar, PSR 1913+16 (Hulse \& Taylor 1975; Burrows 
\& Woosley 1986). In addition, Cyg X-3 is
one of  the   most luminous sources  of   radio  jets in  our  Galaxy,
displaying  repeated  ejections at  (probably) relativistic velocities
(e.g. Schalinski et al.\ 1998).

While much data  have been  gathered  on  the  system through   X-ray,
near-infrared and radio  observations, the most fundamental properties
of the Cyg X-3  system have been difficult   to constrain.  This  is a
consequence     of  line  emission   found  to     be dominated by   a
non-axisymmetric wind component rather than one or both of the stellar
objects.  Furthermore, the wind features are generally weak, very 
broad and at times heavily blended, making line identification 
difficult or impossible. Consequently stellar masses have not been 
measured from binary
motion.  There exist plausible  arguments  for both a high  (Schmutz,
Geballe \&   Schild 1996) and   low (Mitra 1996, 1998)  mass companion
star, which has inevitable consequences  for the state of the  compact
object,  be it a neutron  star  or a black   hole.  As the only  known
Wolf-Rayet + compact  object binary and (probably) the shortest-period
high-mass  X-ray binary,    accurate constraints  on   the  mass   and
dimensions of the Cyg X-3 system are  crucial for our understanding of
many aspects of massive binary evolution.

In a previous paper (Fender, Hanson, \& Pooley 1999, hereafter FHP99),
we found a dramatic spectral change during an outburst of the system.
Though the lines suffer from serious blending, it would appear that during
outburst the spectrum of the 
system evolves from showing high-excitation single-peaked
emission  lines   to one dominated by low-excitation  double-peaked 
profiles.  In this current paper we present a first analysis of these 
data in order to map the velocity field of the line-emitting regions.

\begin{figure*}
\epsfig{file=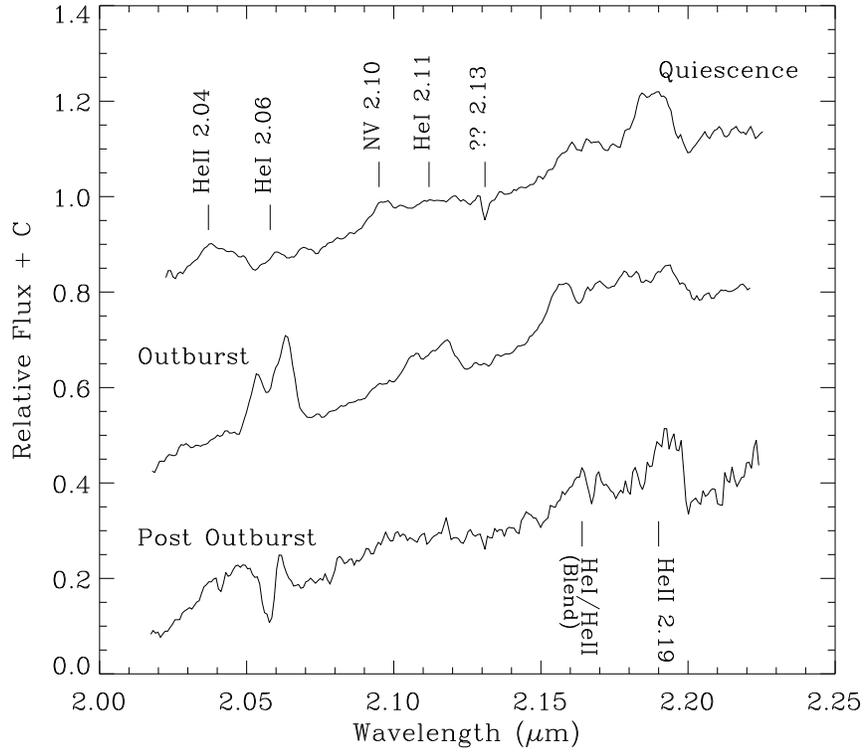, clip=}
\caption{K-band spectra of Cyg X-3, each displayed with an arbitrary
flux offset, `C', for clarity.  The quiescent spectrum shown at top was 
taken 6 June 1996 (epoch A, as defined by FHP99).  The middle spectrum
was taken  19 June 1997 (epoch C)  during very extreme radio and X-ray
flaring of the system.  The bottom spectrum was  taken 15 October 1997
(epoch D). During this time, small X-ray  flaring was still occurring,
but the system was now in `post-outburst'.}
\label{fig1}
\end{figure*}

\section{OBSERVATIONS}

We have recently completed a monitoring  program of the  Cyg X-3 system
spanning 18 months.  Our observations are  
based on broadband  radio observations at 15 GHz combined
with the Rossi  XTE All-Sky Monitor program  (FHP99).  During this
monitoring   program we   obtained  the   best near-infrared
spectroscopy  of the  system,   intermittently throughout the 18 months.
These  data, comprising over 100 individual  spectra taken of Cyg X-3,
were obtained over four different observing campaigns between May 1996
and  October 1997.  The data are  available through  the CDS archive
website (see FHP99).

The near-infrared spectroscopic     data used  in this   analysis  are
presented and fully described  in FHP99. Observations were  made using
the Steward  Observatory's infrared spectrometer,  FSpec (Williams, et
al.\  1993) on the Multiple  Mirror Telescope prior to the recent
upgrade to a single mirror.  The spectra have
an  effective resolution  of R  $\approx$ 1200  at  2.12~$\mu$m, and a
spectral coverage  from 2.02~$\mu$m to 2.22 ~$\mu$m.  Our  analysis in
this paper is based on epochs A, C and D, as defined by FHP99. The Cyg
X-3  system was in    quiescent-,  outburst- and  post-outburst   mode,
respectively, during these three epochs.  
Figure 1  displays spectra taken from each
epoch in order to show the main spectral features present during 
these three epochs.  In quiescence the K-band is dominated by broad,  
weak He~{\sc ii} and  N~{\sc v} emission; in outburst strong twin-peaked 
He~{\sc i}  emission  dominates; for  a  detailed discussion  see
FHP99.

\begin{figure*}
\epsfig{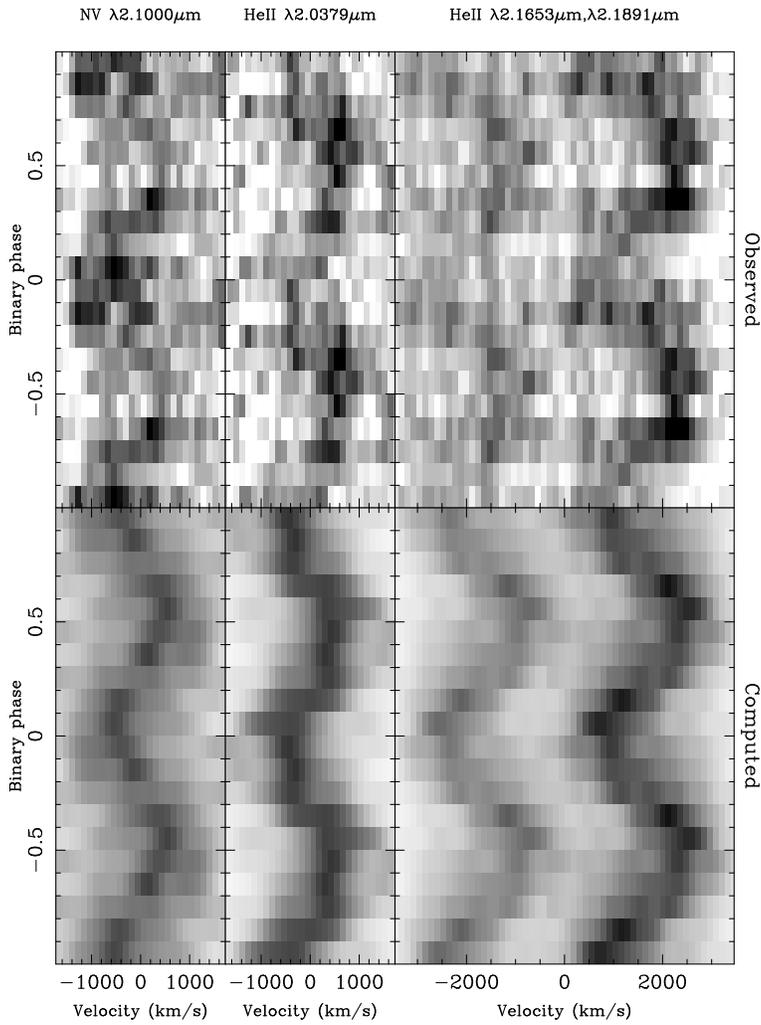}
\caption{K-band trailed spectrograms representing full orbital phase
coverage (4.8   hours) obtained  on  2  June 1996.   Spectral  regions
sampling the four  strongest emission   features  of N~{\sc v} and He~{\sc ii}   are
displayed.  The   data have  been   repeated  over a   second  orbital
cycle.  These data were obtained while  Cyg X-3 was  in quiescence, as
deemed by low radio and X-ray activity (see FHP99).  The lower panel 
shows the model emission profiles produced from the tomographic fit in Fig 3. 
White to black corresponds  to increasing emission intensity on
individual linear scales.  White coincides with the continuum level.}
\label{fig2}
\end{figure*}

\begin{figure*}
\epsfig{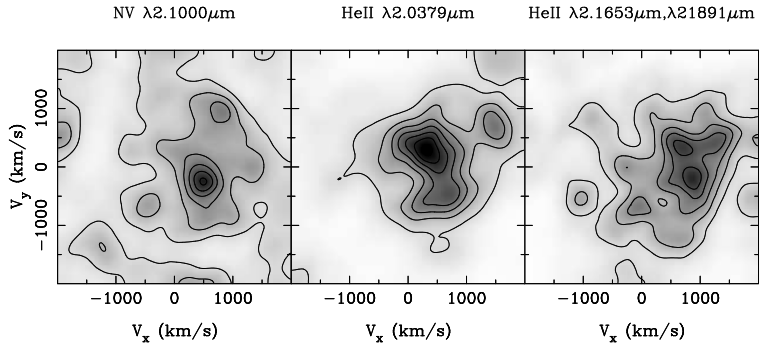}
\caption{Doppler tomograms of the N~{\sc v} and He~{\sc ii} lines during quiescence
in June 1996. These are derived from computed fits of the data 
shown in upper panel of Figure~2. The intensity scales 
are as in Figure~2.  The resulting fit from the tomographic modeling (described in
eqn.\ 1) and their corresponding reduced-$\chi_2$ are given in Table 2.} \label{fig3}
\end{figure*}

\section{ANALYSIS}

Because of the night-to-night variations of the emission line profiles
observed during the outburst phase  (see FHP99), it was important to  
obtain full orbital coverage  on a single night  for this analysis.  We
have time-resolved spectra providing  full orbital sampling during two
nights, 1996 June  2--3, when the  system was  in quiescence and  1997
June 19--20, when   the system was  in outburst.   Despite the heavy
blending, we were able to identify a few of the major features
within the  spectra  using the laboratory  transition
lists of Morris et al.\ (1996).

\subsection{June 1996 Quiescent Spectra}

Trailed spectra from 1996 June are presented in the upper three panels
of Fig 2.  In each case, the  continuum  was removed by  subtracting a
3-spline fit to nearby line-free wavelength  regions and the data were
recast into constant velocity  bins of 121 km\,s$^{-1}$.  Spectra were
further mean-averaged into 10 orbital phase  bins using the ephemeris
of Kitamoto et al.\ (1995).  
Doppler tomograms   were constructed from the  trailed
spectra using the  Maximum Entropy Method  (Marsh and  Horne 1988), an
approach  preferred here  over  other  mapping algorithms because   it
allows  the mapping   of   blended  lines such   as  the  He~{\sc   ii}
2.1653,2.1891~$\mu$m  doublet.   It involves  the  transformation of
velocity--phase information provided by the Doppler-broadened emission
line profiles ($V$,$\Phi$) to a velocity--velocity field ($V_x$,$V_y$)
using:
\[
f(V,\Phi) = \int \int I(V_x,V_y)~~~\times
\]
\begin{equation}
g(V - \gamma + V_x \cos{\Phi} -
V_y \sin{\Phi})~\mbox{d}V_x~\mbox{d}V_y\mbox{.}
\end{equation}
The mapping results from a  $\chi^2$-iteration between model and a fit
and  entropy  minimization  (Skilling and   Bryan 1984).   $f$  is the
emission line intensity at velocity $V$  and orbital phase $\Phi$, $I$
the emission distribution of the resulting map in velocity coordinates
($V_x$,$V_y$), $\gamma$ the systemic  velocity of the binary center of
mass  and  $g(V)$ the local   line profile.  From the combination of
Doppler-broadened line profiles and rotational profile variations over
the orbit, the maps are reconstructions of line emission in the velocity
field around the binary, where $V_y$ is  the velocity in the direction
parallel to a  line joining the  two stellar centers  and $V_x$ is the
velocity orthogonal to $V_y$ in the plane of the orbit.

Any motion  out of  the  orbital plane will not   be recognized by the
mapping and consequently Doppler tomograms  are of most use in systems
where line emission is  confined to gas moving in a plane,  such as
distributed  over an accretion disk or  one of the stellar companions.
If emission originates in a spherical wind the tomograms become difficult, 
if not impossible, to interpret with any confidence.   Our motivation to 
apply tomographic mapping to the lines in Cyg  X-3 is two-fold.  First, in
quiescence, the line emitting  wind region is  thought to be within the
cone   shadowed  from the X-ray   source  by the   companion star (van
Kerkwijk et~al.\  1996).  Motion in the cone is somewhat constrained to
the orbital  plane, although this  assumption breaks  down if the mass
ratio $q =   M_{\mbox{\tiny WR}}/M_{\mbox{\tiny X}}$ is   too large.
Secondly, FHP99 reveal that line emission profiles are double-peaked in
outburst, suggesting that a disk or flattened, rotating wind may dominate
profiles during these epochs, in which case Doppler tomography is well
suited to these spectra.

\begin{deluxetable}{ll|ll}
\tablewidth{0pt}
\tablecaption{$\chi^2_{\nu}$ for Tomographic Fits}
\tablehead{
\multicolumn{2}{c}{Quiescence} &
\multicolumn{2}{c}{Outburst} \\
\colhead{Line ($\mu$m)} &
\colhead{$\chi^2_{\nu}$} &
\colhead{Line ($\mu$m)} &
\colhead{$\chi^2_{\nu}$} 
}
\startdata
N~{\sc v}  2.1000     &   1.9    &   He~{\sc i}  2.0587    &  296.0 \nl
He~{\sc ii} 2.0379   &   5.1    &   He~{\sc i}  2.1126    &   15.0 \nl
He~{\sc ii} 2.1653/2.1891 & 5.2 & He~{\sc i} 2.1623/2.1847 & 31.2 \nl
\enddata
\end{deluxetable}

The tomograms derived from the quiescent data of Fig 2
are shown in Fig 3.  The reduced-$\chi^2$ of these fits
are listed in Table~1.  For reference, the binary center
of mass occurs at ($V_x$,$V_y$) = (0,0) km\,s$^{-1}$,
The companion star center of mass corresponds to
($V_x$,$V_y$)  = (0,$K_{\mbox{\tiny WR}}$), where
$K_{\mbox{\tiny WR}}$ is the radial velocity of the
Wolf-Rayet star, $K_{\mbox{\tiny WR}}$ $\geq 0$. The
compact object center of mass occurs at ($V_x$,$V_y$)  
= (0,$-K_{\mbox{\tiny X}}$) where $K_{\mbox{\tiny X}}$ 
is the radial velocity semi-amplitude of the accreting 
star.  The ratio
$K_{\mbox{\tiny X}}/K_{\mbox{\tiny WR}} = q$, where $q$
is the ratio of stellar masses. Rotating structure
around the compact object or WR star will manifest
itself as an annulus on the map centered on
$K_{\mbox{\tiny X}}$ or $K_{\mbox{\tiny WR}}$,
respectively.  The lower panel of Fig 2 shows the model
emission profiles produced from the tomographic fit in
Fig 3.

\begin{deluxetable}{ll|llll|cc}
\tablewidth{0pt}
\tablecaption{Line Identifications and Measurements, June 1996 Quiescent Spectra}
\tablehead{
\colhead{} &
\colhead{} &
\multicolumn{4}{c}{Gaussian Fit} &
\multicolumn{2}{c}{Tomograms} \\
\colhead{$\lambda_{vac}$} &
\colhead{Line} &
\colhead{$\gamma$} &
\colhead{$K$}  &
\colhead{FWHM}  &
\colhead{$\Phi$}  &
\colhead{$K$}  &
\colhead{$\Phi$} \\
\colhead{($\mu$m)}  &
\colhead{Identification} &
\colhead{(km s$^{-1}$)} &
\colhead{(km s$^{-1}$)} &
\colhead{(km s$^{-1}$)} &
\colhead{} &
\colhead{(km s$^{-1}$)} &
\colhead{} 
}
\startdata
2.0379 & He~{\sc ii} (15-8)&  +162$\pm$28  & 437$\pm$36 & 2959$\pm$201 & $-$0.02$\pm$0.04 & 439$\pm$20 & $-$0.135$\pm$0.05 \nl
2.100  & N~{\sc v}  (11-10)  & $-$108$\pm$22 & 443$\pm$25 & 1928$\pm$322 &  +0.02$\pm$0.02  & 550$\pm$20 & +0.08$\pm$0.04 \nl
2.1653 & He~{\sc ii} (14-8)& $-$ 79$\pm$39 & 446$\pm$26 & 1618$\pm$124 &  +0.00$\pm$0.04  & 890$\pm$20 & +0.03$\pm$0.02 \nl
2.1891 & He~{\sc ii} (10-7)&  +152$\pm$20  & 577$\pm$56 & 2554$\pm$54  &  +0.16$\pm$0.02  &  ''\tablenotemark{a} &   '' \nl  
\enddata
\tablenotetext{a}{Blended calculation}
\end{deluxetable}

\begin{figure*}
\epsfig{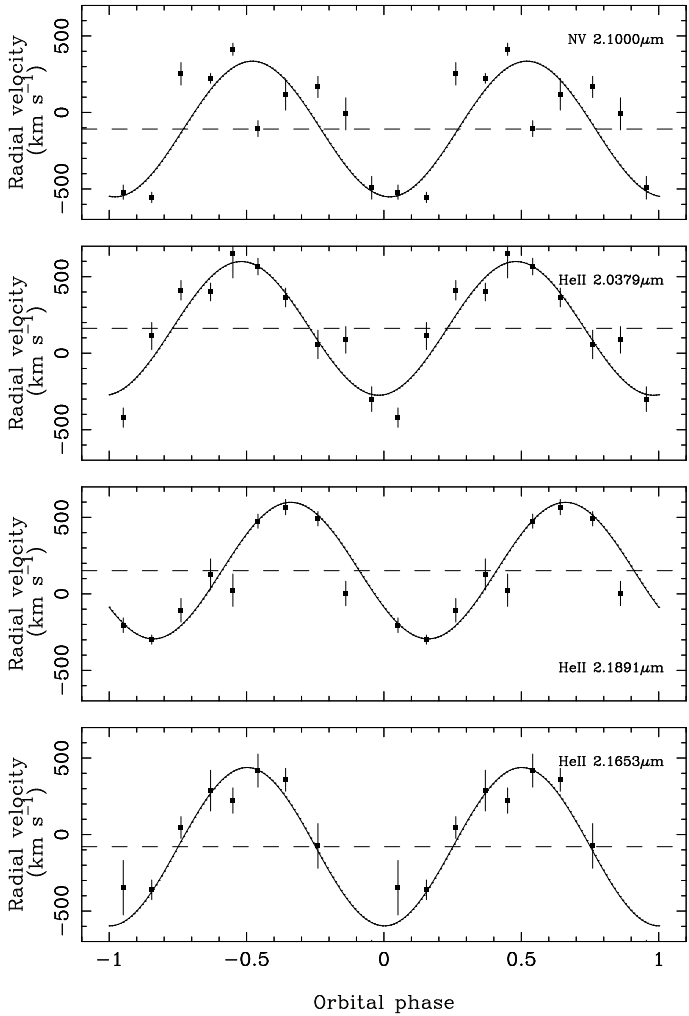}
\figcaption{Radial velocity fits to the four emission   features shown
in Figure 2.  Semi-amplitude    peaks trace general wind   velocities,
which are  also traced in the   tomograms of Figure 3.  The velocities
determined from this and  the tomographic analysis  are given in Table
2.}  \label{fig4}
\end{figure*}

The trailed   spectra and  Doppler tomograms  indicate   that the line
emission  in  quiescence arises in  a  restricted region  of
velocity space. This is consistent with the  van Kerkwijk model in which
the line emission originates in a  region of the stellar  
wind shadowed from ionizing X-rays by the He star.   
Note  that the  cool wind component   and
binary motion vectors are orthogonal: this results in the wind phasing
and WR  star phasing being separated 90 degrees and the tomographic
location of the emission   occurring in a region  offset approximately
$90^{\circ}$  from  the   tomographic   location  of  the   Wolf-Rayet
star. Because of the moderate  $\chi^2$  of these   fits the small scale
structure is not significant.

Two-dimensional Gaussian fits to the tomograms  provide  a first-order
estimate  of both velocity, $K  =  \sqrt{V_x^2 + V_y^2}$, and phasing,
$\Phi =   \tan{^{-1}(V_x/V_y)}$, of  the emitting  gas  and  these are
listed in   Table 2.  Uncertainties   are not propagated   through the
mapping algorithm and therefore the errors are taken to be the size of
a pixel element on the maps.   Similar results are obtained by fitting
one-dimensional, time-dependent Gaussian profiles to the raw  data and
this  provides confidence  in   our  tomographic fits.  These  further
Gaussian  fits   are  also presented  in   Table~2   and the resulting
velocities plotted against orbital phase in Figure~4.

By taking a variance-weighted mean of the one-dimensional fits, the line
forming region(s) have an  average wind velocity  of $K$ = 480 $\pm$ 50
km s$^{-1}$. This is consistent with the value measured by Schmutz et al.\
(1996).  We  do not detect any  significant difference in $K$ 
between the N~{\sc v}
and the three He~{\sc ii} lines in this measure.  From the fits, an average
full-width  half-maximum (FWHM) value  was determined for  each of the
four transitions, and is given in Table 2; if the lines are well represented 
by Gaussian profiles, these values imply a terminal  wind velocity of at
least 1500 km s$^{-1}$ in the system, similar to that found by
van Kerkwijk et al.\ (1996).

We have  also more accurately  determined the  maximum   of  the
blue-shifted  wind   features of  He~{\sc ii}    and  N~{\sc v}  against  the  X-ray
modulation.  From a weighted-mean of  the four measurements, the He~{\sc ii}
and N~{\sc v}  features   are at their  maximum  blue-shift   when  the X-ray
intensity is  at a minimum (e.g.  $\Phi$=0),  with an error of 0.04 in
orbital phase.  

Note  that there  is no well-defined  solution  for the systemic
velocity of the binary center of mass, $\gamma$ (Eqn.\ 1),
from  the 1-dimensional fits in Table 2.  This  could  be the 
result of 
turbulence within the wind (e.g., L\'epine \& Moffat 1999).  
Furthermore, the laboratory wavelength for  the N~{\sc v}   
transition is not accurately known. $\gamma$ can   only   be
constrained to $|\gamma| \leq 200$ km s$^{-1}$, precluding us from
making a distance estimate.  This upper limit on the systemic velocity
seems rather difficult to reconcile with the $+$800 km s$^{-1}$
velocity field recently reported by Chandra observations of the
system (Paerels et al.\ 2000).

\begin{figure*}
\epsfig{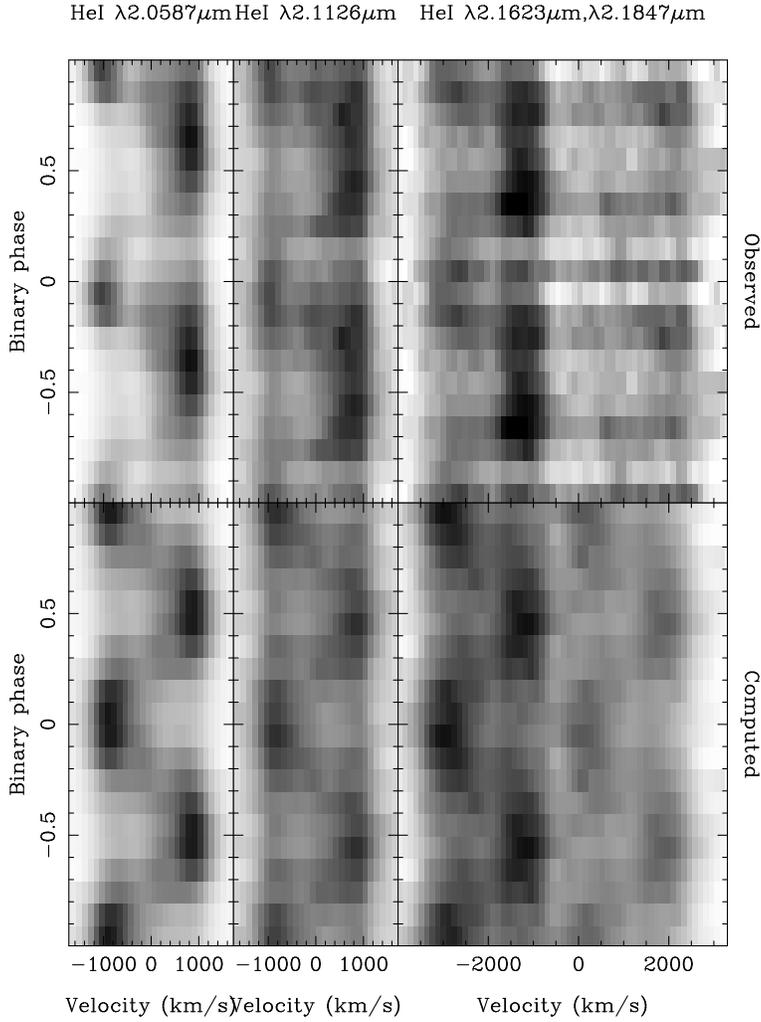}
\caption{K-band trailed spectrograms representing full orbital phase
coverage  (4.8 hours)  obtained  on  19 June   1997.  Spectral regions
sampling   the  four    strongest   emission  features of He~{\sc i} are
displayed. The  data have been repeated over   a second orbital cycle.
These data were obtained while  Cyg X-3 was  in outburst, as deemed by
high radio and X-ray  activity (see FHP99).  The bottom panel shows the
projected fits to the Doppler tomograms given in Figure 6. Similar to Figures
2 and 3,
white to black corresponds  to increasing emission intensity on
individual linear scales, making white coincident with the continuum level.}
\label{fig5}
\end{figure*}

\begin{figure*}
\epsfig{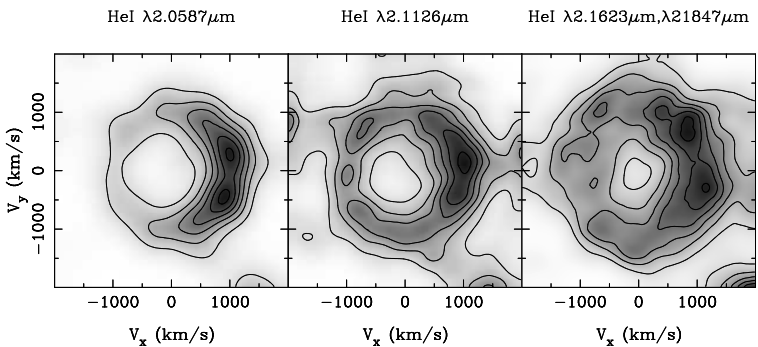}
\caption{Doppler tomograms of the four He~{\sc i} lines during the June 1997 
outburst, derived from the spectra given in the upper panel of Figure 5.  The intensity scales 
are as in Figure~5.} \label{fig6}
\end{figure*}

\begin{figure*}
\epsfig{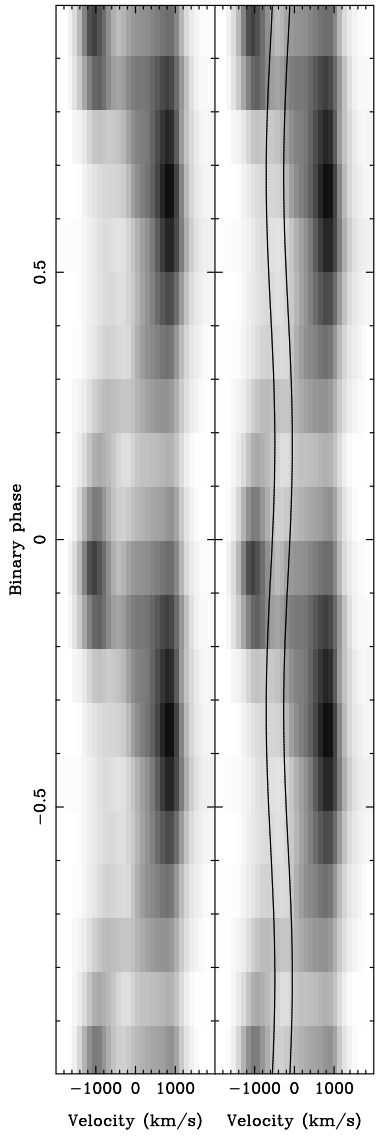}
\caption{An expanded view of the full orbital phase coverage over a 4000
km s$^{-1}$ velocity range centered on the He {\sc i} 2.0587 $\mu$m line, 
as previously shown in Fig.\ 5.  To the right is the same information, though
marked with the location of
the permanently blue-shift absorption feature.  The absorption feature is
seen as white against the broader underlying emission feature of the
2.0587 $\mu$m He {\sc i} line.}  \label{fig7}
\end{figure*}

\subsection{June 1997 Outburst Spectra}

The upper panel of Figure 5 shows the spectra taken during 
the June 1997 outburst.  All
lines seen have been attributed to He~{\sc i}, though exact line
identifications for the $\lambda = 2.16 - 2.18$~$\mu$m region are
difficult to make.  There is likely some contamination from the
2.1891~$\mu$m He~{\sc ii} feature normally seen during quiescence.
Doppler tomograms, derived from the outburst spectra in Figure 5
are shown in Figure 6.   Reduced-$\chi^2$ of each fit are provided in
Table~1. Finally, computed spectra, derived from the tomograms shown 
in Figure 6, are given in the lower panel of Figure 5. 

The line profile behavior of the four He~{\sc i} lines appears at
least marginally similar to that seen during quiescence: strong blue
shifted emission at X-ray minimum and strong red shifted 
emission at X-ray maximum.  However, the 2.0587~$\mu$m  He~{\sc i} line 
in the outburst spectrum
shows much stronger red-shifted emission at $\Phi$ = 0.5 than the
blue shifted emission at $\Phi$ = 0.0 (Fig 5; see also Fig 4 of 
FHP99).  Furthermore, in the region of the the 2.0587~$\mu$m He~{\sc i} 
line, a narrow, permanently blue-shifted absorption feature is
seen, threaded through the background emission (Fig.\ 7).  

The 2.1623~$\mu$m He~{\sc i} line appears almost double-peaked 
though, like the 2.0587~$\mu$m transition, the peak velocities are a 
bit asymmetric, showing strongest emission while at maximum red shift.  
Double-peaked emission may be evidence for rotating gas either 
in an accretion disk or a wind.  However the stronger red peak
also suggests these transitions may be contaminated by a P Cygni
profile. 

All three tomograms describe a ring structure similar to the tomographic
signatures  of  an accretion  disk (Fig.\ 6), consistent with rotational
velocities on the   order  of 1000 km\,s$^{-1}$.    Unfortunately  the
mapping algorithm assumes that  all  line components have a   velocity
modulating symmetrically about $\gamma$.  P Cygni profiles break this
assumption and it  is possible that  the entire ring structure in each
map is the result of P Cygni absorption.  P Cygni components will not be
recreated in the computed spectra and this explains the large discrepancies
between the real data and fits in Fig.\ 5.  It is important that we
consider only the real data in the upper panels when interpreting line
behavior in this system.  Note though that the fits are poor for
the 2.0587~$\mu$m transition (Table 1) but improve  for the other maps
suggesting that  any   P~Cygni   contributions are reduced  in    these
transitions. Consequently the case for a disk or rotating wind cannot be
proven or ruled out in  any of these transitions.

\begin{figure*}
\epsfig{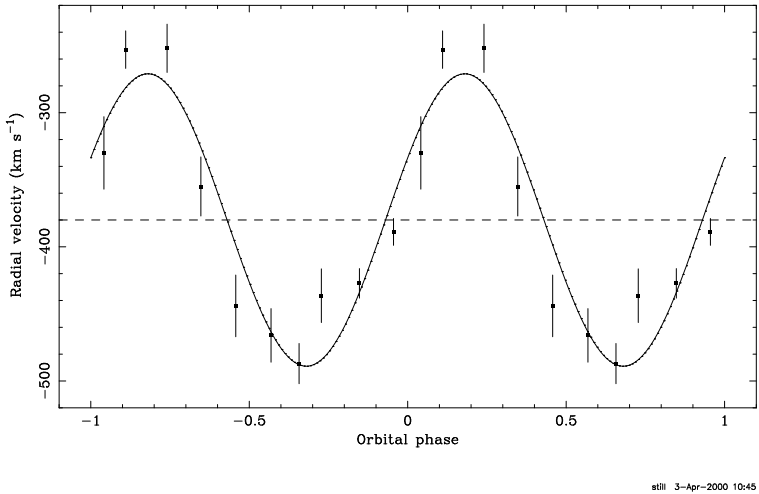}
\caption{Gaussian fit to the absorption trough seen in the
trailed, 2.0587~$\mu$m He~{\sc i} feature in Figure 7, as a function of X-ray
phase.  Parameters   of the fit  include:  $K$   =   109$\pm$13 km  s$^{-1}$,  with   central  velocity
occurring at X-ray phase, $\Phi$ = -0.07$\pm$0.02.  If $\Phi$ is set equal to 0.0,
the corresponding value for $K$, 100$\pm$14 km  s$^{-1}$, is near equally as good
a solution.  The values for $K$ and $\Phi$  are independent of
the exact identification of the spectral feature or features.  The value for 
$\gamma$, shown in the above solution to be -380$\pm$10 km s$^{-1}$, 
assumes the feature is due to the He~{\sc i} 2.0587 $\mu$m line.}  \label{fig8}
\end{figure*}

Fortunately, the sinusoidal absorption feature found in the 2.0587~$\mu$m
spectral region is far less ambiguous, being clearly discerned over the 
entire temporal sequence of Figure 5 and 7.  This continuous  blue
absorption feature was fit with a Gaussian to determine its  velocity
modulation  over the orbital cycle.  Without knowing the shape of the
underlying emission   profile, the fits to the absorption feature are 
dubious.  However
since the  absorption feature is narrow,  being just  a few resolution
elements across, any bias to the result should be small.  The
feature was found to  oscillate with a semi-amplitude  of $K = 109 \pm
13$ km s$^{-1}$.  Most importantly, the phase  of the velocity shifts,
shown in Figure 8, are not like those seen in the N~{\sc v} and He~{\sc ii}
features during quiescence (Figure~2) or like the He~{\sc i} emission features
during outburst, whose red and blue maxima occur at $\Phi$
= 0.5 and 0.0 (Figure 4).  Instead,  the absorption shows red and blue
maximum at phases,  $\Phi$ =  0.18 and 0.68  (with an  error of $\sim$
0.02), relative to the  orbital  ephemeris of Kitamoto et al.\ (1995). 

If $\Phi = 0$ corresponds to superior conjunction of the compact
object in Cyg  X-3, red and blue maxima of features associated with
the mass donor He star should occur at $\Phi = 0.25$ and $\Phi  = 0.75$,
respectively.  Thus the observed  phasing of the absorption feature is
approximately consistent with it tracing the motion of the He star.  The
impact of this result is explored further in  Section 4.1.  Fixing the
phases of the maximum blue and red-shift to correspond with the binary
phases $\Phi = 0.25$ and $\Phi  = 0.75$ and re-fitting  the 
radial velocity curves produces only a slightly  poorer fit, $K = 100 
\pm  14$ km s$^{-1}$, but with a consistent semi-amplitude.

\section{DISCUSSION}

\subsection{The He~{\sc i} 2.0587 $\mu$m transition}

Some very important differences  in the formation of the He~{\sc i}
transitions have been described.  This has lead to very different line fluxes and
quite  possibly different line   profiles seen  in the numerous  lines
detected in the  Cyg   X-3  outburst spectrum.  First,  2.0587~$\mu$m
emission is produced by direct radiative recombination into the 2$^1P$
state, or through recombinations  at higher levels which then cascades
down into  the 2$^1P$  state. This is   not  a problem in   very dense
extended winds,  as the  radiation  field pumps electrons via  the He~{\sc i}
resonance   lines  into upper  levels,  leading   to an enhancement of
recombinations to the 2 $^1P$ level.  However, once in this state, the
atom has only a $\sim$ 1:1000 chance of decaying via the 2.0587~$\mu$m (2
$^1P$ $-$ 2 $^1 S$) transition compared to the 584 \AA\ (2  $^1P$ $-$ 1 $^1S$)
resonant transition.     Thus  strong 2.0587~$\mu$m   emission further
requires a  very high optical depth in  the 584 \AA\  line. There is a
second equally important point about the He~{\sc i} 2.0587~$\mu$m line.  Its
lower  level, 2    $^1 S$, is  meta-stable   and  becomes increasingly
overpopulated (relative to LTE).  Frequently,  a 2.0587~$\mu$m P  Cygni
profile   is seen in  hot,  dense WR  winds   because in the outermost
regions  the  absorption process becomes  quickly favored  due  to the
overpopulated lower level (Williams \& Eenens 1989).  Thus, the 
expected behavior of the 2.0587~$\mu$m line in an extended wind is 
for it to self-absorb readily, i.e., create a P Cygni feature. It is of 
interest to note  that several months after the outburst in June 1997, 
the dominant feature in the spectrum of Cyg X-3
is the P Cygni profile occurring at 2.0587~$\mu$m (Fig 1, see also
FHP99).

The remaining transitions, most specifically the $4s  - 3p$ triplet at
2.1126~$\mu$m (the singlet  $4s - 3p$ line at  2.1138~$\mu$m has about
one third   the  flux  of   the  combined emission  from the   triplet
transition),  are not   directly coupled  to   a resonance transition.
Their emission level    is set only   by  the  recombination   rate of
He$^+$. Furthermore, they are not  predisposed to increased absorption
in dense winds like the 2.0587~$\mu$m transition, since their lower states
readily cascade to lower levels.  These differences can give rise to rather
different behaviors in the He~{\sc i} lines seen in our outburst spectra,
and may possibly explain why the weak sinusoidal absorption feature 
seen threaded through outburst spectrum shown in Figure 5, is detected only
in the 2.0587~$\mu$m He~{\sc i} line.

\begin{figure*}
{\epsfig{file=fig9a.ps, clip=, angle=270, width=10cm}
\quad{\epsfig{file=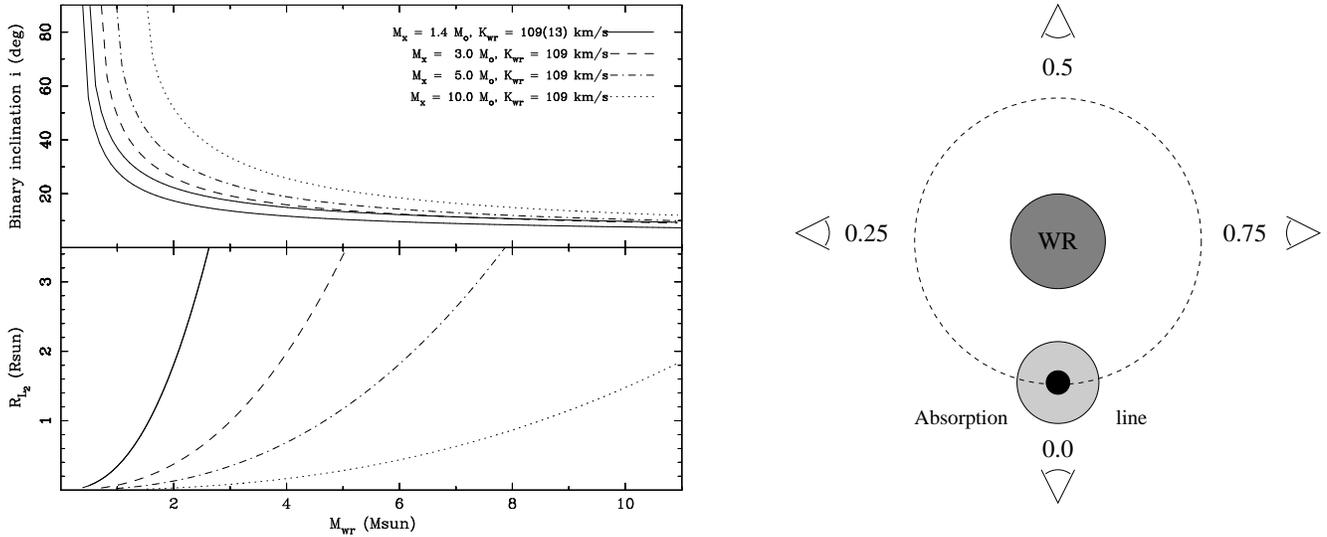, clip=, angle=270, width=7cm}}}
\caption{Orbital solutions assuming the motion traced in Figure 7 and 8 is
centered on or associated with the motion  of the compact object.  
On the left (9a) are shown orbital solutions for differing input
mass of the companion (WR) object,  displayed versus inclination angle
of the binary system and Roche  lobe radius.  Solutions explored range
from  a compact object  mass,  M$_{\mbox{\tiny X}}$ = 1.4 to 10 M$_{\odot}$. 
Two solid curves are given for the  M$_{\mbox{\tiny X}}$ = 1.4 M$_{\odot}$,
solution, which illustrates the plus-minus error range on $K$.
For   any
reasonable WR  star  mass,   M  $>$  2  M$_{\odot}$,  only  extremely  low
inclinations  are possible. This is   difficult to reconcile with  the
X-ray modulation. The
orbital orientation implied by this interpretation is indicated in the
panel to the right (9b).  Such an interpretation
is counterintuitive in the sense that 
it requires X-ray orbital minimum to occur when the X-ray source is nearest
to the observer.  All of these evidences disfavor the interpretation 
that the sinusoidal
variations  are centered on or associated   with motion of the compact
object. }
\label{fig9}
\end{figure*}

\begin{figure*}
{\epsfig{file=fig10a.ps, clip=, angle=270, width=10cm}
\quad{\epsfig{file=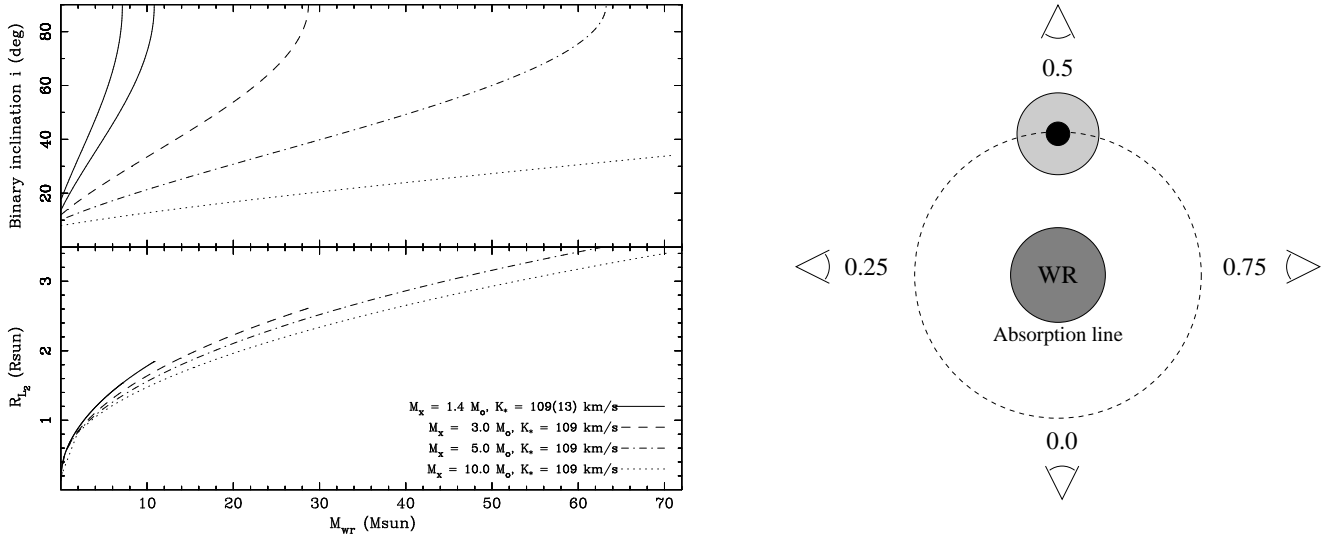, clip=, angle=270, width=7cm}}}
\caption{Orbital solutions assuming the motion traced in Figure 7 and 8 is
centered on or associated with the motion of the mass donor He star. 
On the left (10a) is shown orbital solutions for differing input mass of the
compact object,  M$_{\mbox{\tiny X}}$ = 1.4 to 10 M$_{\odot}$, 
displayed versus inclination and Roche lobe radius.  Two solid curves 
are given for the  M$_{\mbox{\tiny X}}$ = 1.4 M$_{\odot}$,
solution, which illustrates the plus-minus error range on $K$.
This second set of solutions allows for the high inclination suggested
by the X-ray and infrared modulation simultaneously with moderately 
high mass solutions and Roche lobe radii for the mass donor. 
The orbital orientation implied    
by this interpretation is indicated in the panel to the right (10b).
It allows for X-ray orbital minimum to occur when the X-ray source is 
furthest from the observer, at superior conjunction.  This configuration 
gives the most reasonable set of solutions for the orbit of the Cyg X-3
system.  }
\label{fig10}
\end{figure*}

\subsection{Orbital dynamics from the He~{\sc i} 2.0587~$\mu$m absorption line}

Fortunately, an exact identification of the absorption feature seen
in Figure 5 is not necessary for our purposes.  While we feel confident
the feature is due to He~{\sc i} 2.0587~$\mu$m given the arguments 
given above, even if the line is unknown, or due to an unresolved 
blend of features, we can still use its velocity signature given in $K$
to measure the orbital dynamics of the Cyg X-3 system.
As already noted, the phasing of the
absorption feature seems consistent with a source which is distributed
approximately isotropically around the star and hence should reflect
the binary motion of the He star in the system. However, before
continuing with that interpretation, we note that there have been
alternative suggestions for the phasing of the system.  For example
Schmutz et al.\ (1996) propose that the He~{\sc ii} and N~{\sc v}
emission lines commonly observed in quiescence (Figure~2) track the
motion of the He star (in which case X-ray minimum light would be
occurring one quarter cycle  before superior conjunction of the
compact accretor). So it is clear that there is not a consensus 
about the interpretation of the phasing in the Cyg X-3
system.  As a result we have investigated the implications of an
origin for the absorption line to originate in material centered 
on the accretor and on the donor. 

Dynamical solutions for the first case, that the absorption line in the
material is centered on the accretor, are presented in Fig 9a. 
The solid curves show the mass solution assuming the compact object is
a neutron star of M$_{x}$ = 1.4 M$_{\odot}$.  
For this solution only, two solid curves are given to 
demonstrate the error range on the velocity measurement (109 $\pm$ 
13 km s$^{-1}$).  This orientation, where the absorption is centered
on the accretor, is illustrated in Fig 9b.   We see from Fig 9a 
that for all accretor masses, unless the He star is much less massive 
than implied from luminosity arguments (i.e.\  $\leq 2$M$_{\odot}$), the
inclination of the system must be much lower ($\leq 50^{\circ}$) than 
is generally accepted based on the X-ray and infrared modulation. 
Furthermore, any solution that allows for a reasonable inclination,  
$\geq 50^{\circ}$, simultaneously gives an absurdly small, nearly 
non-physical, value for the Roche lobe radius of the donor.  As a 
result we do not consider an origin for the absorption line in the 
accretion disc or any other material centered on the accretor to be 
very likely.

Rather, our analysis implies the most likely configuration is the 
second case, that the material creating the absorption feature is
centered on the mass donor.
Solutions for this interpretation are presented in Fig 10a and
a schematic illustrating the orientation with phase is shown in
Fig 10b. Again, two solid curves are given to illustrate the range
of solutions consistent with the mass accretor being a neutron
star, M$_{x}$ = 1.4 M$_{\odot}$, given our error on the velocity
measurement. In this orientation,  X-ray minimum occurs at superior 
conjunction of  the accretor.  It is immediately apparent from Fig 10a 
that realistic high-inclination solutions are now available.  Using 
this interpretation we directly  derive  the mass  function of  the  
system, which  is 0.027~M$_{\odot}$.  The reason this does not 
constrain at all the
nature of the  accretor is the large, probably  dominant, mass  of the
He donor star.  For a  neutron star accretor of 1.4  M$_{\odot}$
the mass  of the He star   is   most  likely  in  the range
 5~M$_{\odot}$
$\la$ M$_{\mbox{\tiny WR}}$ $\la$ 11~M$_{\odot}$, assuming the
system has a moderately high inclination, $i > 60^{\circ}$.

However, if  the   accretor is  a   black hole,   there is  much  less
confidence about the mass of  the He star.  Charles (1998) lists
derived  masses for black holes in   {\em low-mass} X-ray binaries and
finds them to be in the range 5--12 M$_{\odot}$. Careful inspection of
Fig 10a reveals that  this allows  a wide  range of  possible solutions,
most   of which are    compatible   with very   large  masses   indeed
(i.e.\ several tens of M$_{\odot}$) for the He star.  This is because
the Roche radius given in Figure 10a increases for greater He star mass
and easily stays within the expected radius limit for such stars (Langer 
1989).  Moreover, the Roche 
radius of the He star shown in Figure 10a can even be less than its 
R$_\star$, as shown in Langer's Figure 5.  This is because the tenuous 
outer atmospheric radius of the star, identified as R$_\star$,  
contributes an insignificant fraction of the mass of the star, and 
can extend well beyond the dynamical Roche radius.  Nonetheless, we 
see from Figure 10a that solutions for a M$_{\mbox{\tiny BH}}$ $\ga$ 
10 M$_{\odot}$ would
be inconsistent with our assumption of even a moderate inclination 
for the binary system, for any realistic companion star mass, 
M$_{\mbox{\tiny WR}}$ $\la$ 70 M$_{\odot}$.  For this reason, we put 
an upper limit on the mass of the compact object of M$_{\mbox{\tiny BH}}$ 
$\la$ 10 M$_{\odot}$.  The well known high mass X-ray binary system 
Cyg X-1 has a similar mass estimate for its black hole (Charles 1998).

We  note that  in  the  binary   orientation  proposed  by  Schmutz et
al.\ (1996), which lead them to argue for a most likely value of 
M$_{\mbox{\tiny BH}}$ $\approx$ 17 M$_{\odot}$ for the compact object, 
the absorption  feature would be  required to arise from  a region 
centered not on the He star or  accretor, but a quarter of a cycle  
out of phase with either orientation.

\subsection{The nature of the stellar companion}

The assumptions used   to determine stellar   mass in  Figure 10a also
provide limits  for the   radius  of the   stellar
companion of Cyg X-3.   We are able to test  whether this result is
consistent with the characteristics, most specifically the luminosity,
predicted from atmosphere models and observations of typical WR stars.

The distance  to Cyg  X-3 is  not directly known.   However, Dickey
(1983) has suggested a  lower  limit on  the distance, 11.6   $\times$
$\pi$/10 kpc, where $\pi$ is the distance in kpc to the galactic center,
 based on  the detection of high  velocity neutral
hydrogen  absorption features and applying  a  flat rotation curve for
the  Galaxy.  With an  updated value for  the distance to the Galactic
center of $\sim$~8 kpc (Reid  1993), this gives a  lower limit to  Cyg
X-3   of  $\sim$~9 kpc.   Original estimates  of   the  line of  sight
extinction to Cyg  X-3, A$_V=15$,  were  made  by Becklin et   al.\
(1973).  More recent measurements  suggest a slightly higher  value of
A$_V=19$ (Molnar et al.\ 1988), A$_J=5.5$  (van Kerkwijk et al.\ 1996)
which corresponds approximately  to A$_V=19$, and A$_J=6.0$ (Fender et
al.\ 1996)  corresponding    to approximately   A$_V=20.5$ (Rieke   \&
Lebofsky 1985).  Determining the line of sight extinction is dependent
on   measuring the flux  distribution  over a  large enough wavelength
range to note the differential change, as done by both van Kerkwijk et
al.\ and Fender et al.  But equally important, one must simultaneously
make assumptions about the functional form  of that extinction and the
functional form  of the underlying continuum  of the source.  For this
reason, without an  {\it  a priori} quantitative understanding  of the
stellar companion,  the line of  sight  extinction cannot be as well
constrained as one might hope.  Mindful of these  limitations, we have
used the apparent magnitudes  of Cyg X-3   given by Wagner  et al.\
(1989) and Fender et al.\ (1996), $m_I = 20.0$,  $m_H = 13.11$, $m_K =
11.76$ and applied an extinction  of  approximately $A_V= 20$   and  a
distance  of  9 kpc to obtain  absolute  magnitudes of  $M_I = -4.8$,
$M_H=-5.2$, $M_K=-5.1$ for the stellar companion.  It is precisely
these values, the very luminous lower limits for the absolute magnitude 
of the stellar companion, that most strongly challenge the notion of 
Cyg X-3 being a low mass X-ray binary system (Mitra 1998)

We have argued that while in quiescence, the emission lines seen in Cyg
X-3, presumably dominated by the companion  star, look most similar to
the lines seen in early WN-type (WNE) WR stars (FHP99). 
Based on direct  measurements  of WNE stars   given in  Smith  et al.\
(1994), a mass range from 6 to 12~M$_{\odot}$, corresponds to a typical,
 $-4 > M_V > -4.8$.   Naturally, there  is a large  scatter in the
observed relation.     Without  a   direct determination   of   $M_V$,
with the uncertainties in the bolometric correction and with the difficulty
in extrapolating the extinction characteristics  towards Cyg X-3, the
luminosity for the stellar     component of Cyg   X-3 is   not  highly
inconsistent with the luminosity of other WNE stars.

\subsection{The nature of the accretor}

The crucial question of whether the accretor is a neutron star
or a black hole cannot be decided  with confidence. The only {\em
direct} observational evidence for the presence of a neutron star in
Cyg X-3 was in the report of 12.6 ms $\gamma$-ray pulsations, $\sim
15$ years ago (Chadwick et al.\ 1985).  However, this result has never
been widely accepted.  As already stated, Schmutz et al.\ (1996) argue
that the observational evidence supports a black hole accretor, which
prompted Ergma \& Yungelson (1998) to investigate the evolution of
such a system.  They concluded that Cyg X-3 {\em may} contain a black
hole accreting at super-Eddington rates.  Mitra (1998) however argued strongly
{\em against} the interpretation of Schmutz et al.\ (1996), preferring
a low-mass system (which is extremely hard to reconcile with
the very high luminosity of the system).  For now we will have to
consider a fairly large range of accretor masses up to 
M$_{\mbox{\tiny BH}}$ $\la$ 10 M$_{\odot}$ (see discussion in \S 4.2).

\section{CONCLUSIONS}

The X-ray binary, Cyg X-3, is the only known system
containing a (presumably) massive He star with a compact
object.  This makes Cyg X-3 a uniquely important link in
our understanding and testing of massive X-ray binary
evolution.  In this paper we pursued an emission line
analysis of the quiescent and outburst spectra presented
in FHP99.  We reported that while the double-peaked
emission seen in the He~{\sc i} lines during outburst are
consistent with a disk-wind geometry as proposed in
FHP99, our tomographic analysis in particular reveals
unresolvable ambiguities in the mechanism responsible
for the formation of double-peaked emission.  For this
reason, additional constraining spectral information or
methods are needed to resolve the nature of the
double-peaked emission and its relation to the wind
or disk geometry of the system during outburst.

Of greater significance was the detection of a weak
absorption feature threading through the blue wing of
the 2.0587~$\mu$m He~{\sc i} line.  This 
feature moves $\sim$ 1/4 out of phase with all other
spectral modulations seen in Cyg X-3, consistent with
He~{\sc i} absorption originating in an isotropic, asymptotic
wind from the companion star.  Consequently it can be
used to derive the first radial velocity curve for the
Cyg X-3 system.  Employing reasonable geometric
assumptions, we derive a mass range of 5~M$_{\odot}$
$\la$ M$_{\mbox{\tiny WR}}$ $\la$ 11~M$_{\odot}$ for the
He star, if the accretor is a neutron star.
Additionally, should the accretor be a black hole, we
determine M$_{\mbox{\tiny BH}}$ $\la$ 10 M$_{\odot}$,
based on an upper mass limit, M $\la$ 70 M$_{\odot}$,
for the He star companion.

Identification of the critical absorption feature used
in obtaining a mass function for the Cyg X-3 system was
made possible only through phased-resolved near-infrared
observations.  A further key was obtaining observations
during outburst.  Clearly, periods of high X-ray
variability would make the most opportune times to catch
this rare infrared state once more.  The spectra presented
in FHP99 represent the best quality a 4-meter class
telescope will likely ever deliver on this object, and yet
these spectra barely detect the very weak, sometimes
very broad, and often heavily blended features.  Its 
not certain that moving to an 8-meter class telescope
would resolve all the problems with the current data set, 
as many of the lines are already resolved with the 
current spectra.  However, the narrow absorption 
feature detected in the outburst spectrum may be just
the first of many similarly weak and narrow features which
might better be revealed with higher-quality spectra.  Such
features offer promise in resolving the ambiguities
plaguing our analysis.

FHP99 showed there to be
a marked decrease in X-ray flux during the four days
prior to the near-infrared outburst, the minimum
occurring just after outburst.  This may have resulted
from drops in the ionization state of the wind following a
temporary decrease in the X-ray flux.  FHP99 showed that
the outburst occurring in June 1997 and previous
outbursts suggested in earlier spectra (van Kerkwijk et 
al.\ 1996) were
short lived ($\leq$ 24-h).  Careful day-to-day
monitoring of Cyg X-3 in the X-ray and radio,
particularly during epochs of high X-ray and radio
activity, would potentially precede suitable windows for
near-infrared spectroscopic observations of P Cygni lines.

\acknowledgements

We are grateful for comments made by our referee which lead us to
greatly improve the presentation of this work.
Observations reported in this paper were obtained with the Multiple Mirror 
Telescope, operated by the Smithsonian Astrophysical Observatory and the 
University of Arizona. We are grateful to G.\ and M.\ Rieke for their critical
support in this program. M. M. H. received support for this work provided by
The University of Cincinnati through a University Research Council Grant and 
a Faculty Summer Fellowship.

\end{document}